\documentclass[useAMS,usenatbib,fleqn]{mn2e}

\usepackage{times}

\usepackage{deluxetable}
\usepackage{graphicx}
\usepackage{epsfig}
\usepackage{amssymb,amsmath}
\usepackage{aas_macros}
\usepackage{fixltx2e} 
\title[Submillimetre wavelength Sgr A* variability]{An 8 hour characteristic time-scale in submillimetre light curves of Sagittarius A*}
\author[Dexter et al.]{Jason Dexter$^{1}$\thanks{E-mail:
    jdexter@berkeley.edu}, Brandon Kelly$^{2}$, Geoffrey C. Bower$^{1}$, Daniel P. Marrone$^{3}$,\newauthor Jordan Stone$^{3}$, Richard Plambeck$^{1}$, and Sheperd S. Doeleman$^{4,5}$\\
$^{1}$Department of Astronomy, Hearst Field Annex, University of California, Berkeley, CA 94720-3411, USA\\
$^{2}$Department of Physics, Broida Hall, University of California, Santa Barbara, CA 93106-9530, USA\\
$^{3}$Steward Observatory, Arizona Radio Observatory, University of Arizona, 933 North Cherry Avenue, Tucson, AZ 85721-0065, USA\\
$^{4}$Massachusetts Institute of Technology, Haystack Observatory, Route 40, Westford, MA 01886, USA\\
$^{5}$Harvard-Smithsonian Center for Astrophysics, 60 Garden St., Cambridge, MA 02138, USA}

\begin{document}

\pagerange{\pageref{firstpage}--\pageref{lastpage}} \pubyear{2013}
\maketitle

\label{firstpage}

\begin{abstract}
We compile and analyse long term ($\approx 10$ year) submillimetre ($1.3$, $0.87$, $0.43$ mm, submm) wavelength light curves of the Galactic centre black hole, Sagittarius A*. The $0.87$ and $0.43$ mm data are taken from the literature, while the majority of the $1.3$ mm light curve is from previously unpublished SMA and CARMA data. We use Monte Carlo simulations to show that on minute to few hour time-scales the variability is consistent with a red noise process with a $230$ GHz power spectrum slope of $\beta=2.3^{+0.8}_{-0.6}$ at $95\%$ confidence. The light curve is de-correlated (white noise) on very long (month to year) times. In order to identify the transition time between red and white noise, we model the light curves as a stochastic damped random walk process. The models allow a quantitative estimate of this physical characteristic time-scale of $\tau = 8_{-4}^{+3}$ hours at 230 GHz at $95\%$ confidence, with consistent results at $345$ and $690$ GHz. This corresponds to $\approx 10$ orbital times or $\approx 1$ inflow (viscous) time at $R = 3 R_s$, a typical radius producing the 230 GHz emission as measured by very long baseline interferometry and found in theoretical accretion flow and jet models. This time-scale is significantly shorter (longer) than those measured at radio (near-infrared) wavelengths, and is marginally inconsistent with the same variability mechanism operating in the submm and NIR for the expected $t \propto R^{3/2}$ scaling. It is crudely consistent with the analogous time-scale inferred in studies of quasar optical light curves after accounting for the difference in emission radius. We find evidence that the submm variability persists at least down to the ISCO, if not the event horizon. These results can be compared quantitatively with similar analyses at different wavebands to test for connections between the variability mechanisms, and with light curves from theoretical models of accreting black holes. 
\end{abstract}


\begin{figure*}
\includegraphics[scale=.6]{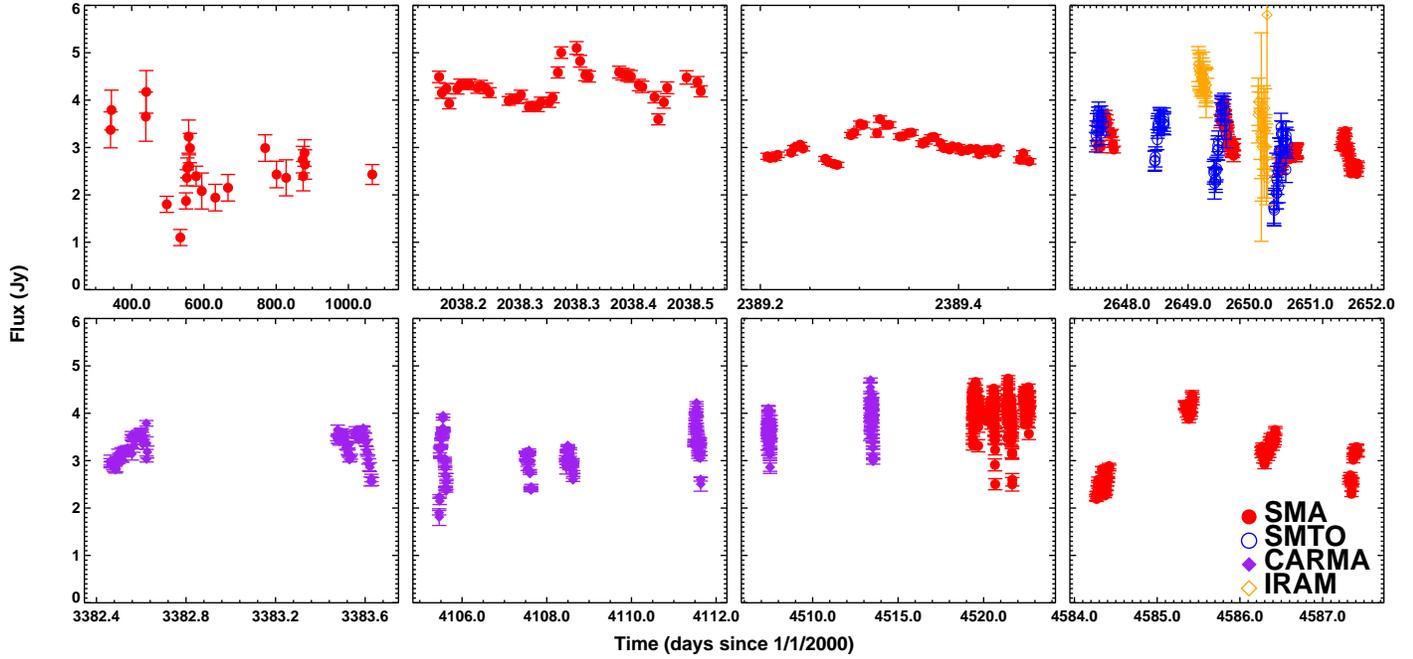}
\caption{\label{lc230}Full Sgr A* 230 GHz light curve. Data points are taken from (panels in order from top left): \citet{zhao2003}, \citet[][middle two panels]{marrone2008}, and \citet{yusefzadehetal2009}. The bottom four panels are previously unpublished CARMA and SMA data. Note the irregular sampling.}
\end{figure*}

\begin{figure*}
\includegraphics[scale=.6]{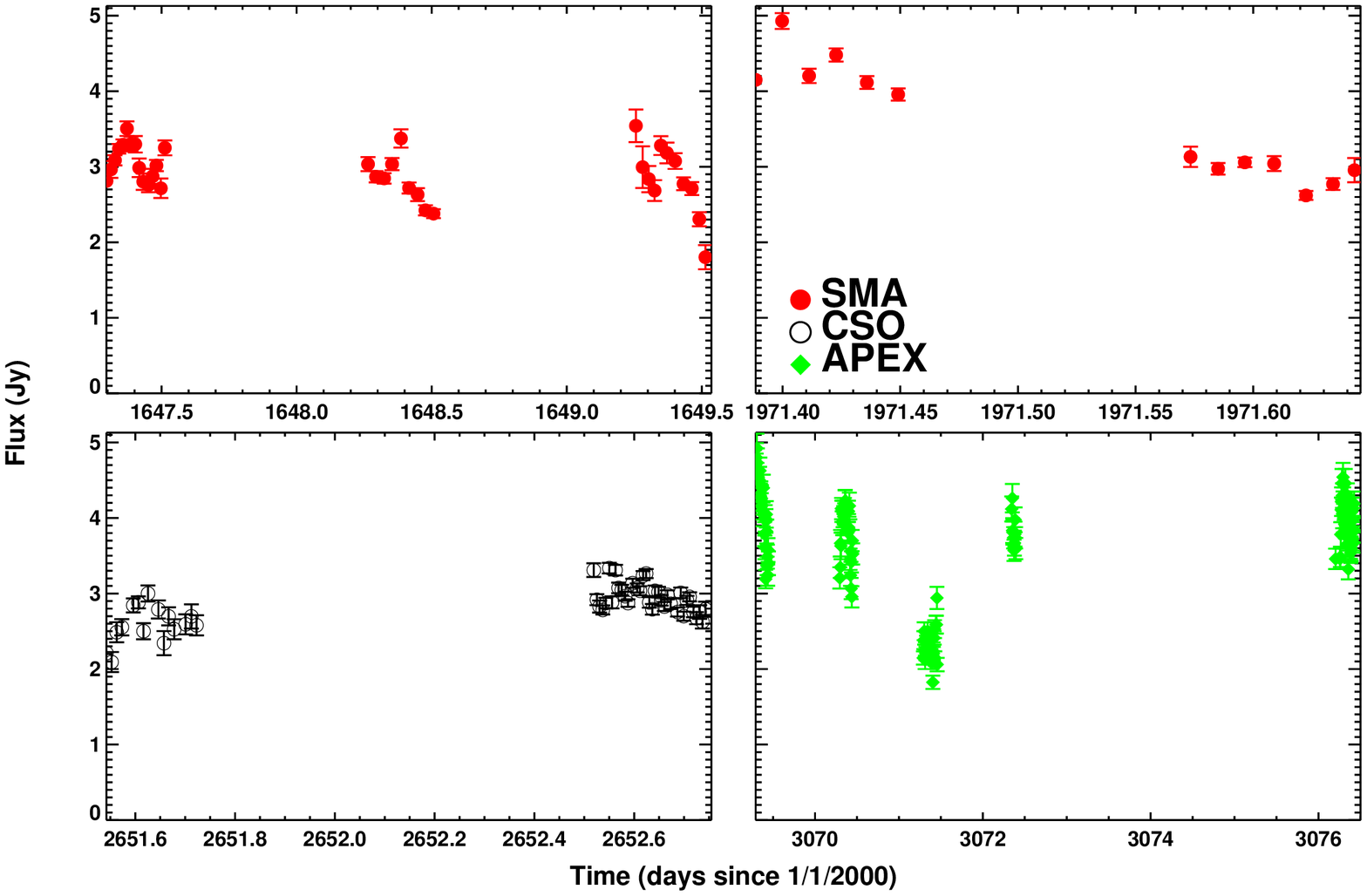}
\caption{\label{lc345}Full Sgr A* 345 GHz light curve. Data are taken from \citet[][SMA, top left panel]{eckartetal2006}, \citet[][SMA, top right panel]{marroneetal2006}, \citet[][CSO, bottom left panel]{yusefzadehetal2009}, and \citet[][APEX, bottom right panel]{garciamarinetal2011}.}
\end{figure*}

\begin{figure*}
\includegraphics[scale=.6]{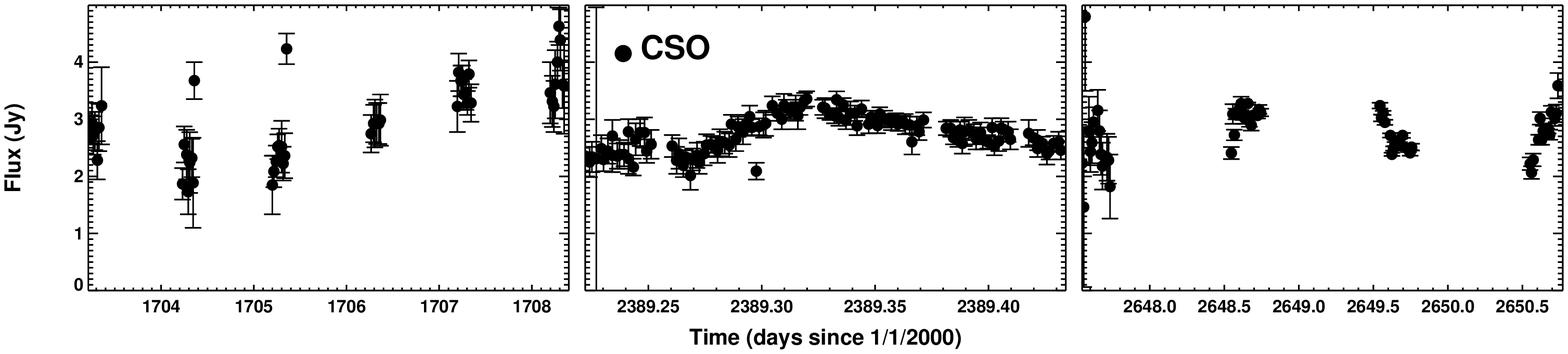}
\caption{\label{lc690}Full Sgr A* 690 GHz light curve. The data are from the CSO, taken from \citet{yusefzadehetal2008,yusefzadehetal2009}.}
\end{figure*}

\section{Introduction}

The Galactic centre black hole, Sagittarius A* (Sgr A*) is the most intensively studied supermassive black hole candidate. Discovered as a compact radio source \citep{balick1974}, its spectrum rises to a peak in the submillimetre \citep[``submm bump'',][]{zylkaetal1995}. The properties of the submm bump are particularly well constrained. The observed Faraday rotation and linear polarization give strong constraints on the density of the emitting region \citep[][]{bower03,marrone2007}, while very long baseline interferometry (VLBI) observations have determined the size of the emitting region up to $1.3$ mm  \citep[e.g.,][]{lo98,krich98,shen05,bower2006,doeleman2008} and the temperature of the emitting electrons from the brightness temperature. 

Long term monitoring in the radio over the past decade has established a steady characteristic flux density of $\simeq 3$ Jy in the submm, with characteristic fluctuations of $30-50\%$ on hour time-scales \citep{zhao2003,eckart2008sim,marrone2008}. Sgr A* has also been observed to exhibit large amplitude, rapid ($\simeq 30$ min) variability in the near-infrared \citep{genzel2003,ghez2004} and X-ray \citep{baganoff2001} bands. The X-ray ``flares'' occur simultaneously with peaks in the near-infrared \citep{eckartetal2004,doddseden2009}. Several multi-wavelength campaigns have looked for connections between the flares and radio to submm emission by cross-correlating single epoch light curves between different wavebands. Some detections of increased submm and radio flux density following NIR/X-ray activity have been reported, and in some cases these detections have been interpreted as emission from hot plasma cooling by adiabatic expansion \citep{yusefzadeh2006,marrone2008,eckartetal2012}. In other cases, anti-correlations have been found, and these events have been interpreted as decreases in magnetic field strength from reconnection \citep{doddsedenetal2010,hauboisetal2012} or absorption from intervening material \citep{yusefzadehetal2010}. There is no consensus as to the physical connections between radio, submm, and NIR/X-ray emission, or whether there is a connection at all.

All of the observations of Sgr A* have led to long, irregularly sampled light curves over the past decade or so. Statistical analyses of individual light curves have shown that at least some of the radio variability must be intrinsic to Sgr A* rather than the effect of interstellar scintillation \citep{macquartbower2006}, and have shown that the NIR emission can be described as a red noise process \citep{doetal2009} with a non-linear flux distribution \citep{doddsedenetal2011,witzeletal2012} and a power spectrum break at $\simeq 150$ min \citep{meyeretal2009}. 

Here, we compile submm\footnote{The light curves have average observed wavelengths of $1.3$, $0.87$, and $0.43$ mm. We refer to them collectively as submm for simplicity and to distinguish from previous studies at $\ge 3$ mm.} light curves of Sgr A* from 2001 to 2012, using a combination of published and unpublished data from several telescopes (described in \S\ref{data}). We study the statistical properties (power spectra) of the light curves from individual days (\S\ref{lightcurves}), and find the variability to be consistent with a stationary noise process on minute to hour time-scales. At very long time separations, we expect the variability to be uncorrelated. We tentatively identify a turnover in the structure function from red to white noise. However, the structure function and other non-parametric methods for studying variability are biased when used on irregular light curves. Similar structure functions have been found in optical quasar variability studies \citep{hughesetal1992,vandenberketal2004}, and recently characteristic variability amplitudes and time-scales have been estimated from these light curves by modeling the process as a damped random walk \citep{kellyetal2009,kozlowskietal2010,macleod2010}. This method is described and used to estimate parameters of our light curves in \S\ref{drw}. The implications of these results for physical models of Sgr A* are discussed in \S\ref{discuss}.

\section{Observational data}
\label{data}

We compile light curves from a combination of published and unpublished data from several telescopes at $230$, $345$, and $690$ GHz. Previously published data are from the SMA \citep{zhao2003,eckartetal2006,marroneetal2006,marrone2008,yusefzadehetal2009}, CSO \citep{yusefzadehetal2008,yusefzadehetal2009}, SMTO \citep{yusefzadehetal2009}, IRAM 30m telescope \citep{yusefzadehetal2009}, and APEX \citep{eckart2008sim,garciamarinetal2011}. The previously unpublished light curves from CARMA and SMA are discussed below.

\subsection{CARMA}

The CARMA 230 GHz flux densities were measured during mm-VLBI campaigns in 2009 Apr \citep{fishetal2011} and 2011 Mar, and as a byproduct of polarization observations in 2012 May.  An integration time of 10 sec was used.  System temperatures, scaled to outside the atmosphere, were calibrated every 10 minutes using a chopper wheel.  Observations of secondary flux calibrators (1751+096, 1733-130, 1924-292) interleaved with the Sgr A* scans every 10 to 15 minutes were used to monitor gain variations of the antennas.  The flux density scale was established using observations of Uranus (assuming a disk brightness temperature of 102 K) and MWC349 (assuming a flux density of 1.95 Jy); the absolute flux scale is estimated to be uncertain by $\pm 15\%$.  Sgr A* flux densities were measured using only interferometer spacings $> 20 k\lambda$, to resolve out contaminating emission from the mini-spiral and other large scale structures in the Galactic centre.  The total duration of the Sgr A* observations was limited to approximately 4 hours per day, the time the source is above 18 degrees elevation at CARMA. The integrations were averaged in 5 minute intervals, and the errors were estimated from the scatter within each interval. 

\subsection{SMA}

The SMA monitored Sgr A* for 8 days in 2012, divided into four-day runs in May and July. During each run, seven antennas were used. In May they were arranged in the compact configuration (baseline lengths of 7--52 k$\lambda$), and in July the antennas were arranged in the subcompact configuration (4-19 k$\lambda$). Observations of the source 1733-130 were interleaved with observations of Sgr A* in order to track changes in the complex gain. Less frequent observations of 1743-038 and 1924-292 were used to check that the gain solutions were accurate. The absolute flux density scale was determined by observations of planets (Uranus, Neptune, Titan), or through comparison with calibrator fluxes in the SMA database \citep{gurwelletal2007}. The scale is uncertain by $15\%$. 

Typically, the light curve of SgrA* is derived from baselines longer than $20$ k$\lambda$, as in the CARMA data described above, because of confusion with extended structure surrounding Sgr A* on shorter baselines. However, in the case of the July data, no baselines are longer than $20$ k$\lambda$. We employed an iterative procedure to remove point source emission from short time intervals in each day of data and construct a common SgrA*-free map of the surrounding material across all four days in each month. To begin, for each timestep over each four-day run we fit a central point source to the visibilities and subtract it. The point-source-subtracted visibilities from each day of the four-day runs are then consolidated, inverted, and CLEANed to produce a first-guess image of the unrelated emission. These images are converted to visibilities and subtracted from the data from each night. These new visibilities, in which most of the surrounding emission has been eliminated, are subjected to the same procedure -- after removing point sources at each timestamp a new image is created from the four days and subtracted from the visibilities. This is iterated many times, resulting in an image of the mini-spiral as well as a Sgr A* light curve with the mini-spiral contribution removed. Since the July data were obtained without long baselines, contaminating emission from within a few arcseconds of SgrA* cannot be distinguished from the central source. To correct for this effect, the visibility functions of the May and July minispiral maps were matched where they overlap in uv space by allowing a central offset component and a scale difference between the two. This procedure shows that $0.6$ Jy of unresolved contamination was present in the July light curves. 

\subsection{Light curves}

The full light curves at $230$, $345$, and $690$ GHz are shown in Figures \ref{lc230}-\ref{lc690}.\footnote{A table of the previously unpublished $230$ GHz light curve data are available as an online supplement to this article.} Large gaps have been removed, and the different symbols denote data from different telescopes. Significantly more data are available at $230$ GHz, so that this light curve ultimately is the most useful for measuring variability properties. The entirety of the $345$ and $690$ GHz light curves are taken from the literature, while the majority of the $230$ GHz light curve (all of the CARMA data and the majority of the SMA data) is new here (bottom four panels in Figure \ref{lc230}). There are 1044, 245, and 225 data points at $230$/$345$/$690$ GHz spanning 29, 11, and 14 epochs of observations. 

In the following sections, we study the variability properties of these light curves over the full range of minute to year time intervals using a few different techniques.

\begin{figure}
\includegraphics[scale=.8]{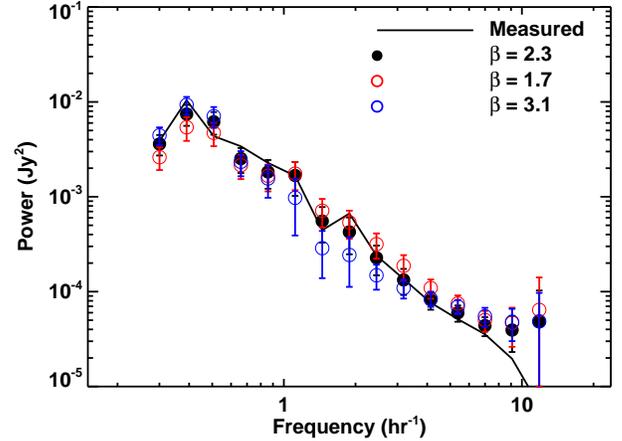}
\caption{\label{psd}Observed (solid line) 230 GHz power spectrum combining all individual observing days as well as the best power law fit (filled circles) and the maximum and minimum allowed slopes at $95\%$ confidence (open circles). The model power spectra are strongly distorted by windowing effects from the irregular sampling patterns and instrumental errors in the observations.}
\end{figure}

\begin{figure}
\includegraphics[scale=.8]{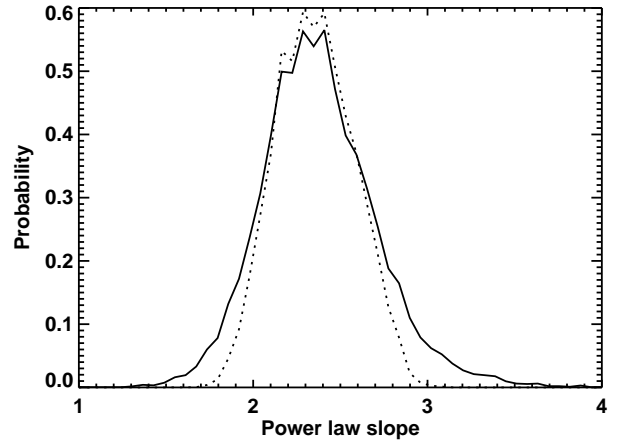}
\caption{\label{rejprob}Probability that the power law model for various slopes provides an acceptable fit to the 230 GHz power spectrum, as estimated directly from the distribution of simulated power spectra (solid line) and from $\chi^2$ statistics (dashed line).}
\end{figure}

\section{Intraday power spectrum}
\label{lightcurves}

We first consider the intraday variability on times ranging from the separation between individual integrations ($\gtrsim 5$ minutes) to the total duration of the light curves from individual telescopes ($\lesssim 5$ hours). In general, the light curves even for individual days are irregularly sampled and contain gaps. Smoothed power spectra from each day are calculated from regularly sampled, continuous light curves as follows. We re-bin each day's light curve to a common interval, and linearly interpolate to fill in empty bins caused by larger gaps. The new time bin is taken as the largest sampling interval present in the data which is smaller than 3 times the most commonly used sampling interval. This is an arbitrary choice but gives good frequency coverage for the power spectra. The resulting regular light curve is Fourier transformed to calculate the power spectrum. Then this power spectrum is binned logarithmically with factors of $1.3$ between frequency bins \citep{papadakislawrence1993}. 

The resulting individual day power spectra can be characterized as single power laws with slopes $P(f) \propto f^{-\beta}$ in the range $\beta = 0.5-3$. However, both red noise leak and aliasing effects resulting from the re-binning and interpolating procedure can strongly bias raw power spectra calculated in this way. In order to quantitatively measure the power law slope, we follow \citet{uttleyetal2002} and simulate long, continuous random light curves from power law power spectrum models of varying slopes using the method of \citet{timmerkoenig1995}. We then sample the random light curves at the same times as the observations from a given day, mimic instrumental errors by adding Gaussian noise at the same level as the errors at each observed time, and use the same re-binning / interpolating procedure described above to produce continuous light curves in the same way as the data. We use $3000$ simulated light curves from each model, but note that in most cases converged results at $95\%$ confidence only require $300$.

The feasibility of the model can be assessed by fitting the mean of the simulated power spectra to the observed one, and comparing the residuals to the dispersion in the simulated light curves. Formally, we estimate a $\chi^2$ value from:

\begin{equation}
\chi^2 = \sum_i \frac{(P_\nu - \langle P_{\nu, \rm sim}\rangle)^2}{{\Delta P_{\nu, \rm sim}}^2},
\end{equation}

\noindent where $\langle P_{\nu, \rm sim}\rangle$ is the mean of the simulated power spectra and the errors, $\Delta P_{\nu, \rm sim}$, are their rms dispersion in each frequency bin. The distribution of the $\chi^2$ values calculated between the individual realizations of the simulated power spectra and $\langle P_{\nu, \rm sim} \rangle$ allows us to evaluate the quality of the fit between the model and data. The power law model provides a satisfactory fit at $95\%$ confidence to all of the days in all light curves, and the allowed ranges of the power law slope overlap between almost all days ($26/29$ days for $230$ GHz, all for $345$ and $690$ GHz). This suggests that i) the power law model can describe the individual day light curves, ii) each light curve is independently not very constraining and ii) there is no evidence that the variability process is non-stationary. 

In order to better constrain the power law slope, we then apply the same power law model to all of the days. The same procedure is followed as above, except that the unbinned power spectra from each day are combined and then logarithmically binned to form the observed intraday power spectrum for the entire light curve. We fit models to this power spectrum again as above, binning simulated power spectra from all days, fitting them to the observed one, and calculating the allowed range of slope.

The observed $230$ GHz power spectrum is shown in Figure \ref{psd} (solid line), along with the best-fitting power law model (filled circles) and the models with the smallest and larger allowed slopes at $95\%$ confidence (open circles). The model fits each have different error bars, corresponding to the rms error at each frequency calculated from the full range of simulated light curves. The constraint on the power law slope from this procedure is $\beta=2.3^{+0.8}_{-0.6}$ at $95\%$ confidence, and the probability as a function of $\beta$ that the power law models give an acceptable fit to the data is shown in Figure \ref{rejprob}. The model power spectra are highly distorted by the re-binning and interpolating procedure. A naive fit to the raw power spectrum, ignoring the effects of irregular sampling and observational errors, gives abest-fitting power law of slope $\beta \simeq 1.8$, $2 \sigma$ away from the actual best model. This demonstrates the importance of accounting for windowing effects and observational noise. The slope constraints at $345$ and $690$ GHz are $2.1_{-1.1}^{+1.7}$ and $2.0_{-0.4}^{+0.8}$ ($95\%$ confidence). The $690$ GHz featureless power law spectra are highly distorted by instrumental errors in the light curves and irregular sampling, so that the best-fitting model is only marginally compatible with the data from all days combined. If this model can in fact describe the data, the allowed range of power law slope is likely larger than formally estimated.

We can further look for power at particular frequencies that is in excess of what is expected for ourbest-fitting models. At $99.7\%$ confidence, both the binned and unbinned power spectra from all days combined fall within the range present in the simulated power law models. This suggests that if there is excess power, from a quasi-periodic oscillation (QPO) or some other source, it is below the sensitivity of these light curves and/or is  transient. 

How sensitive are our light curves to QPOs? We test this by adding sinusoidal features to the observed $230$ GHz light curve with a range of amplitudes and quality factors ($\nu / \Delta \nu$ where $\Delta \nu$ is the Lorentzian width of the QPO). Between $\nu = 0.4-4 \rm hr^{-1}$ with $\Delta \nu = 0.0 - 0.2 \nu$, excess power appears at $\nu$ or $2 \nu$ at $99.7\%$ confidence for amplitudes of $0.1-0.3$ Jy. The light curve is most sensitive to high quality, short period QPOs.

\begin{figure*}
\begin{center}
\begin{tabular}{lll}
\includegraphics[scale=.6]{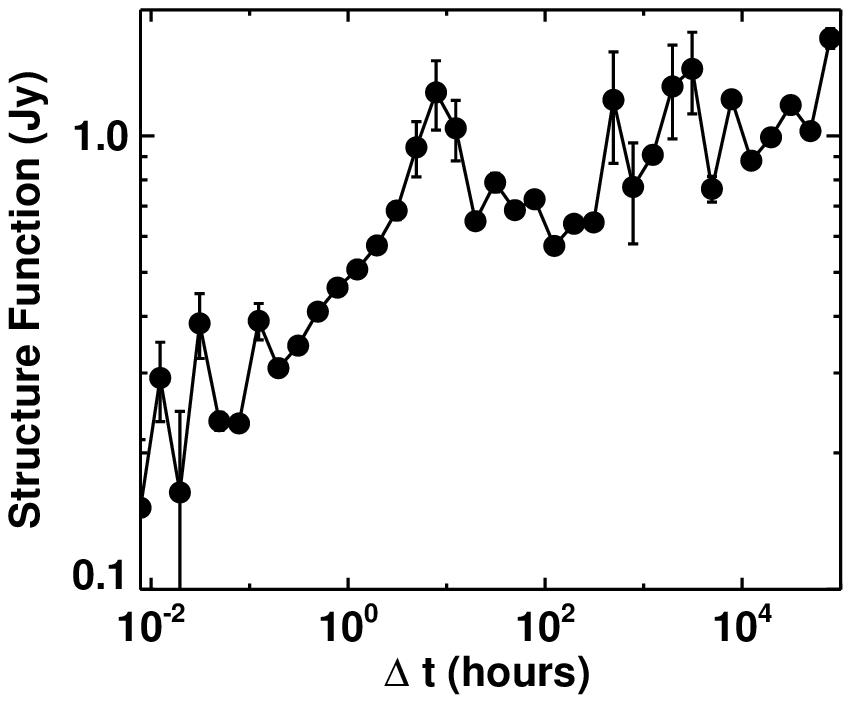} & \includegraphics[scale=.6]{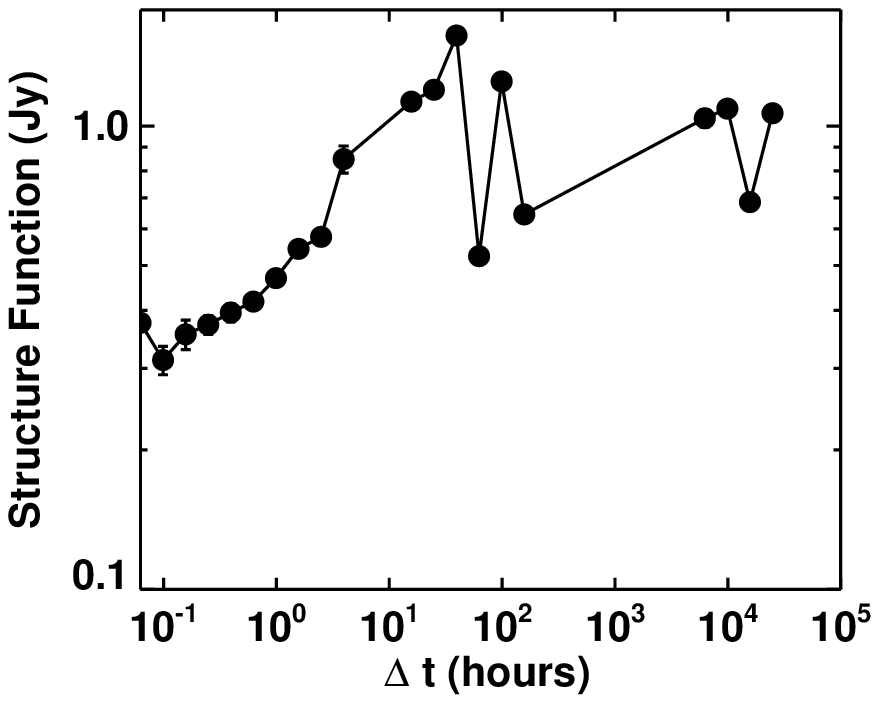}  \includegraphics[scale=.6]{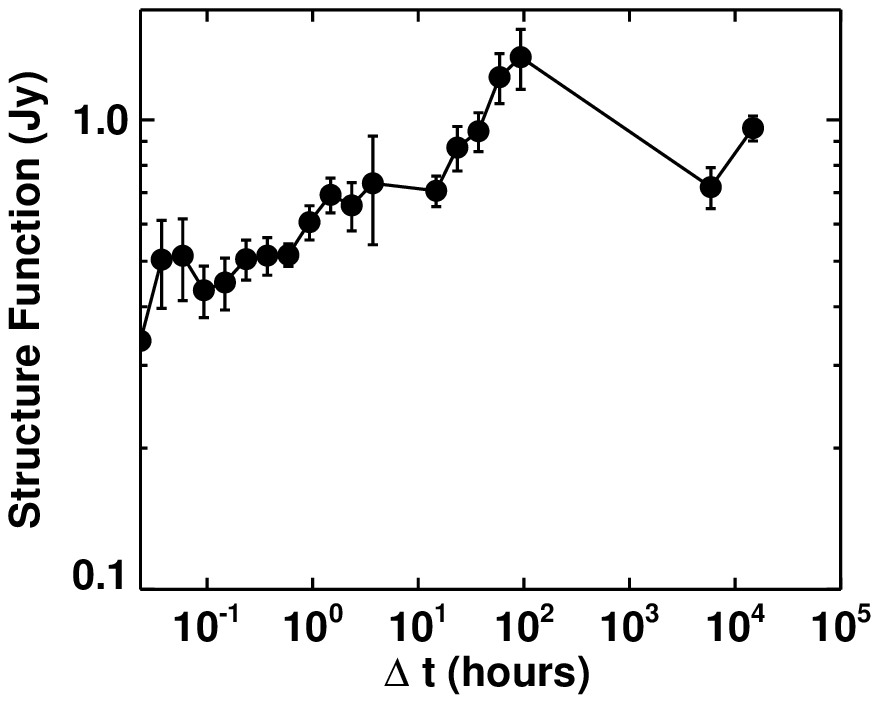}
\end{tabular}
\end{center}
\caption{\label{sf}Structure functions of the light curves at 230, 345, and 690 GHz, binned regularly in $\log{\Delta t}$. The error bars are given by the combined statistical errors, $SF(\Delta t)/\sqrt{N}$, where $N$ is the number of points in each bin, and an estimate of the intrinsic errors by taking the standard deviation of 1000 structure functions calculated from adding Gaussian noise to each light curve point.}
\end{figure*}

\begin{table*}
\caption{Damped Random Walk Parameter Estimates\label{params}}
\begin{small}
\begin{center}
\begin{tabular}{lcccccccc}
        \tableline
 & $\tau$ (hr) & & & & $\hat{\sigma} \hspace{2pt} (\rm Jy / \rm day^{1/2})$ & & \\
Light Curve & Best Value & $1 \sigma$ & $2 \sigma$ & $3 \sigma$ & Best Value & $1 \sigma$ & $2 \sigma$ & $3 \sigma$\\
        \tableline
                  230 GHz&     7.5 &(     6.0,     9.0) &(     4.1,    10.8) &(     3.6,    14.6) &     1.9 &(     1.8,     1.9) &(     1.8,     2.0) &(     1.7,     2.1)\\
              230 GHz log&     9.8 &(     6.2,    14.6) &(     5.1,    18.8) &(     4.6,    23.7) &     1.7 &(     1.7,     1.8) &(     1.6,     1.8) &(     1.6,     1.9)\\
230 GHz $\Delta t > 2$ hr&     8.7 &(     5.6,    11.9) &(     1.4,    16.1) &(     1.0,    24.7) &     1.8 &(     1.5,     2.0) &(     1.2,     2.3) &(     1.1,     2.9)\\
                  345 GHz&     5.6 &(     3.4,     7.8) &(     1.7,    13.5) &(     1.7,    39.3) &     2.2 &(     2.0,     2.3) &(     1.8,     2.5) &(     1.6,     2.7)\\
              345 GHz log&     5.9 &(     2.5,    11.2) &(     2.0,    22.0) &(     1.8,    53.6) &     1.9 &(     1.7,     2.2) &(     1.7,     2.4) &(     1.6,     2.5)\\
                  690 GHz&     6.6 &(     3.2,    10.0) &(     1.2,    24.1) &(     1.2,  3722.2) &     1.5 &(     1.3,     1.7) &(     1.1,     1.9) &(     0.9,     2.1)\\
              690 GHz log&     5.5 &(     1.9,    11.7) &(     1.3,    27.2) &(     1.1,   342.4) &     1.7 &(     1.5,     2.0) &(     1.5,     2.2) &(     1.4,     2.5)\\
                 Combined&     6.4 &(     5.4,     7.4) &(     4.3,     8.5) &(     3.5,    10.4) &     1.9 &(     1.8,     1.9) &(     1.8,     2.0) &(     1.7,     2.1)\\
	\tableline
\end{tabular}
\end{center}
\end{small}
\end{table*}

\section{Full light curve analysis}
\label{drw}

We next consider the variability properties of the full light curves. We first use the structure function, a method for characterizing irregularly sampled light curves which has been used in a number of quasar variability studies \citep[e.g.,]{simonettietal1985,hughesetal1992,vandenberketal2004}. We tentatively identify a characteristic time-scale in the structure function (\S\ref{sec:structure-function}) where the variability transitions from red (correlated) to white (uncorrelated) noise, which we then verify and quantitatively measure by modeling the full  light curves as the result of a stochastic damped random walk noise process (\S\ref{sec:light-curve-modeling}).

\subsection{Structure function}
\label{sec:structure-function}

The first order structure function is defined as,

\begin{equation}
  SF^2 (\Delta t) = \frac{1}{T} \int dt \left[F(t) - F(t + \Delta t)\right]^2,
\end{equation}

\noindent where $F(t)$ here is the light curve (flux density as a function of time) with total duration $T$. For discrete data, the integral becomes a sum over pairs of flux densities separated by $\Delta t$. In practice, this is calculated by summing the squared flux density differences ($\Delta F_j$) over all pairs of time lags present in the light curve, binning by time lag ($\Delta t_i$), and dividing by the number of points in each bin ($N (\Delta t)$):

\begin{equation}
  SF^2 (\Delta t_i) = \frac{1}{N(\Delta t_i)} \sum_j \Delta F_j (\Delta t_i)
\end{equation}

The structure functions of the full light curves at each frequency are shown in Figure \ref{sf}. The submm light curves show significant intrinsic variability, as has long been known \citep[e.g.,][]{zhao2003}. The flux density differences grow from some minimum on short times (set by observational noise) as roughly a power law (consistent with our quantitative power spectrum analysis above) until a turnover at $\sim 1-100$ hours, beyond which they appear roughly constant within the errors ($\rm SF_\infty \simeq 1$ Jy, where $\rm SF_\infty$ is the asymptotic rms power on long time intervals). The error bars are largest on time separations between observing days ($\simeq 10$ hours) and especially between epochs ($\gtrsim 100$ hours). The form of these structure functions is close to that expected for a noise process, where the turnover time is a characteristic, physical scale in the intrinsic variability \citep{hughesetal1992}. This type of structure function is also commonly seen from optical quasar light curves. 

However, structure functions have been shown to produce both spurious and inaccurate bends or breaks for irregularly sampled, featureless simulated data with gaps \citep{emmanoulopoulosetal2010}. In order to confirm that the turnover from red to white noise is not caused by biases in the structure function, and to quantitatively measure its value, we model the light curves directly as a damped random walk.

\subsection{Light curve modeling}
\label{sec:light-curve-modeling}

Recent studies of quasar variability have modeled light curves directly with a stochastic noise process \citep{kellyetal2009,kozlowskietal2010,macleod2010}. This is an alternative to the approach used to study the intraday variability in \S\ref{lightcurves} based on \citet{uttleyetal2002}. This particular CAR(1) (damped random walk, DRW) process gives structure functions similar in shape to that shown in Figure \ref{sf}, where the variability appears to decline as red noise ($\rm SF \sim t^{1/2}$) (saturate as white noise) on short (long) time separations. Modeling light curves in this fashion is free of observational and sampling bias. Given the approximately red noise nature of the intraday variability found from Monte Carlo simulations (\S\ref{lightcurves}) and the white noise properties on long time separations, we expect that this model can accurately describe the variability in the Sgr A* light curves.

Quantitatively, the structure function of the DRW is given as \citep{macleod2010}:

\begin{equation}
  \rm SF_{\rm drw} = \rm SF_{\rm \infty} \left(1-e^{-|\Delta t|/\tau}\right)^{1/2},
\end{equation}

\noindent where $\tau$ is the transition time-scale and $\rm SF_{\infty}$ is the asymptotic power in the light curve on times $\Delta t \gg \tau$.

\subsubsection{Fitting Procedure}

We assume this model can describe the submm Sgr A* light curves, and use it to infer the characteristic parameters $\hat{\sigma} \equiv \rm SF_{\rm \infty}/\sqrt{\tau}$ and $\tau$ as follows. Following \citet{kellyetal2009}, the likelihood of a DRW model fitting the light curve with times $t_i$, flux densities (or alternatively their logarithm) $x_i$, and errors $\sigma_i$ sampled at points $i=1$,...,$N$ is given by a product of Gaussians describing the residuals from each light curve point \citep[c.f.][Eqs. 6-12]{kellyetal2009}:

\begin{align}\label{eq:1}
&p({x} | b, \tau, \hat{\sigma}) = \prod_{i=1}^n \left[2\pi (\Omega_i+\sigma_i^2)\right]^{-1/2} \exp\left(-\chi_i^2/2\right),\\
&\chi_i = \frac{\left(\hat{x}_i - x_i^*\right)^2}{\Omega_i+\sigma_i^2},\\
&\hat{x_i} = a_i x_{i-1} + \frac{a_i \Omega_{i-1}}{\Omega_{i-1}+\sigma_{i-1}^2}\left(x_{i-i}^* - \hat{x}_{i-1}\right),\\
&\Omega_i = \Omega_1 (1-a_i^2) + a_i^2 \Omega_{i-1} \left(1-\frac{\Omega_{i-1}}{\Omega_{i-1}+\sigma_{i-1}^2}\right),\\
&a_i = e^{\left(t_i - t_{i-1}\right)/\tau},
\end{align}

\noindent where $x_i^* = x_i - b\tau$ is the difference between the measured value and its mean ($b \tau$). The DRW excursion from the mean ($\hat{x}_i$) and its variance ($\Omega_i$) are calculated recursively from initial values $\hat{x}_1=0$, $\Omega_1=\tau \hat{\sigma}^2 / 2$. The residual between the DRW fit value and the actual light curve is then given by $\chi_i$, where the factor $\Omega_i + \sigma_i^2$ is the sum of the variance of the DRW expected at time $i$ and that from the observational errors in the light curve. The DRW variance ($\Omega_i$) is equal to the long time variance ($\Omega_1$) if neighboring time points are separated by $\Delta t \gg \tau$ ($a=0$), and approaches zero with decreasing time separation ($a=1$) for perfect data ($\sigma_i^2=0$). In the intermediate range, $\Omega_i$ is given by a linear combination of these limits depending on the expected correlation between points ($a_i$).

We can convert the above likelihood of the observed light curves arising from a given set of DRW parameters into the posterior probability of the set of model parameters using a Bayesian approach. Following \citet{kellyetal2009}, we assume uniform priors on $\log{\tau}$, $b$, and $\hat{\sigma}$. The fit quality can be assessed by noting that the residuals, $\chi_i$, for the best-fitting parameters should be independent and normally distributed. Therefore, a Gaussian fit to a histogram of its residuals should have a mean of zero and a variance of unity.

Random samples of $b$, $\hat{\sigma}$, and $\tau$ from the DRW are drawn using a Metropolis-Hastings algorithm. Parameters are estimated as the median value of each parameter sampled, while allowed ranges at a given significance are estimated as the minimum and maximum parameter values found within the region containing the desired fraction of the total probability density.

\subsubsection{Parameter Estimates}

\begin{figure}
\includegraphics[scale=.85]{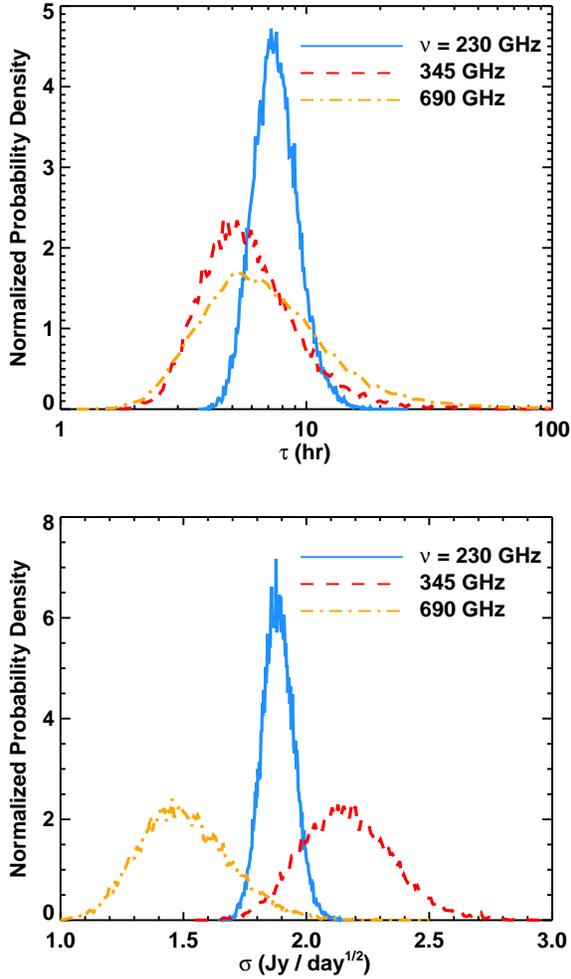}
\caption{\label{tausig230}Posterior probability distributions in each of the DRW model parameters for each of the observed light curves. Characteristic timescales of several hours are found in all cases.}
\end{figure}

The probability distributions over $\tau$ and $\hat{\sigma}$ from fitting the DRW model to the Sgr A* submm light curves at each observed frequency are shown in Figure \ref{tausig230}. The inferred parameter values to $95\%$ confidence are $\tau=8_{-4}^{+3}$ hours and $\hat{\sigma}=1.9 \pm 0.1$ (Jy / $\sqrt{\rm day}$) at $230$ GHz. The parameter $\hat{\sigma}$ measures the stochastic variability power on a given time interval on the red noise portion of the power spectrum. The inferred $\tau$ and $\hat{\sigma}$ values for the $345$ and $690$ GHz light curves are consistent with the results at $230$ GHz at $95\%$ confidence. If we assume that the variability properties are identical at all frequencies, then we can get better statistics by combining the light curves. We re-scale the light curves so that the mean flux densities are equal. To prevent unphysical correlations in time, we offset the times between different frequencies so that they are separated by $\Delta t = 100 \hspace{2pt} \rm days \gg \tau$. Fitting the combined data the allowed $95\%$ confidence range in $\hat{\sigma}$ is unchanged, while the time-scale becomes $\tau = 6^{+4}_{-2}$ hours, a tighter constraint than using $230$ GHz alone but consistent with that range. However, it is possible that this estimate is biased due to the marginal difference in allowed $\hat{\sigma}$ between the light curves. In particular, this may bias the fits towards smaller $\tau$ values.

\citet{kellyetal2009} argued that fitting the logarithm of the flux density is more appropriate than the flux density itself, because the DRW can generate both positive and negative values while the flux density is strictly positive. We find consistent results when using $\log F_\nu$. As discussed below, the fit quality is similar in both cases. We conclude that the data cannot distinguish between a stochastic Gaussian process in $F_\nu$ or $\log F_\nu$. The latter was suggested by \citet{uttleyetal2005} for X-ray variability in AGN. 

\begin{figure*}
\begin{center}
\begin{tabular}{lll}
\includegraphics[scale=.56]{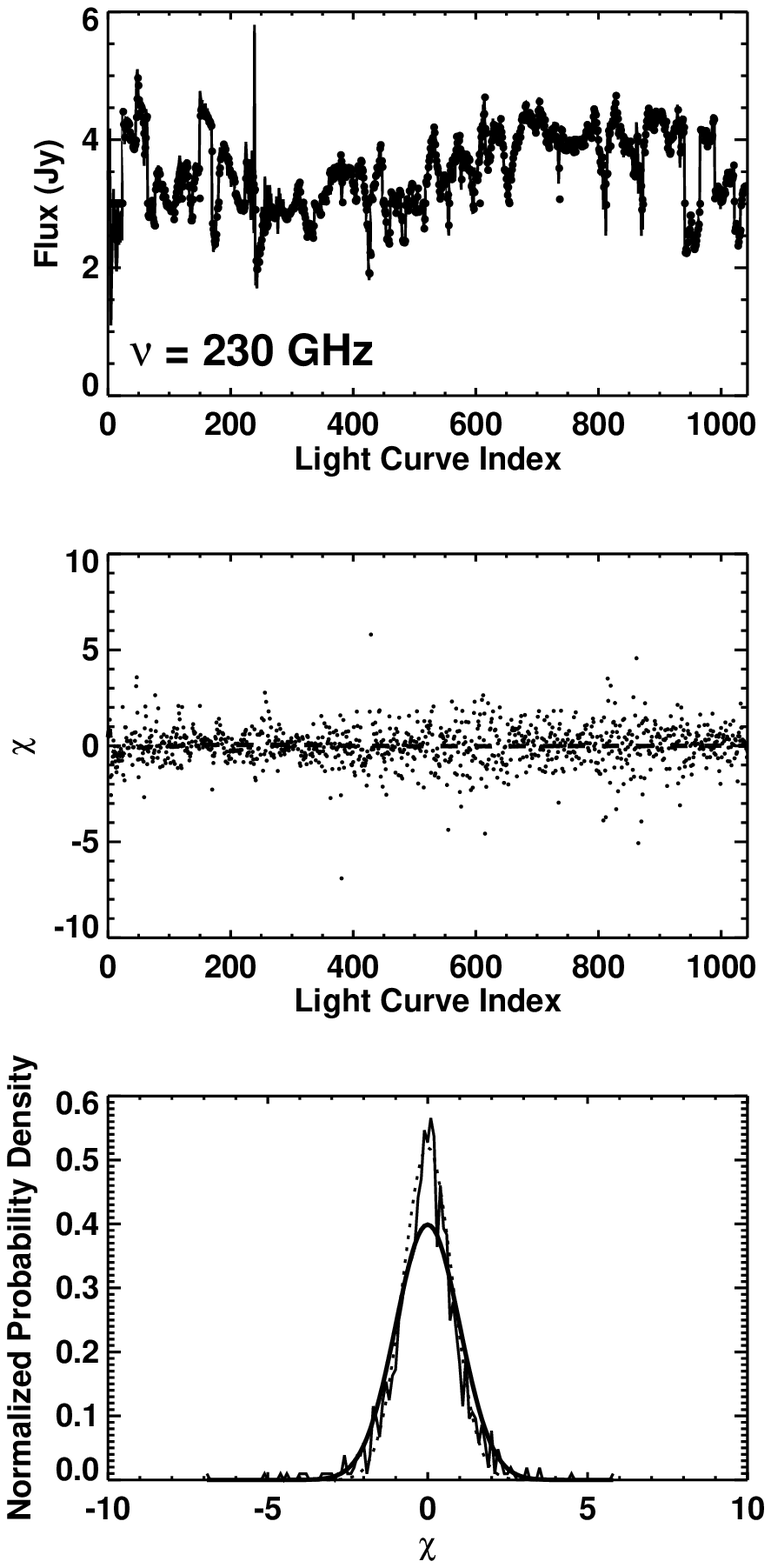} &
\includegraphics[scale=.56]{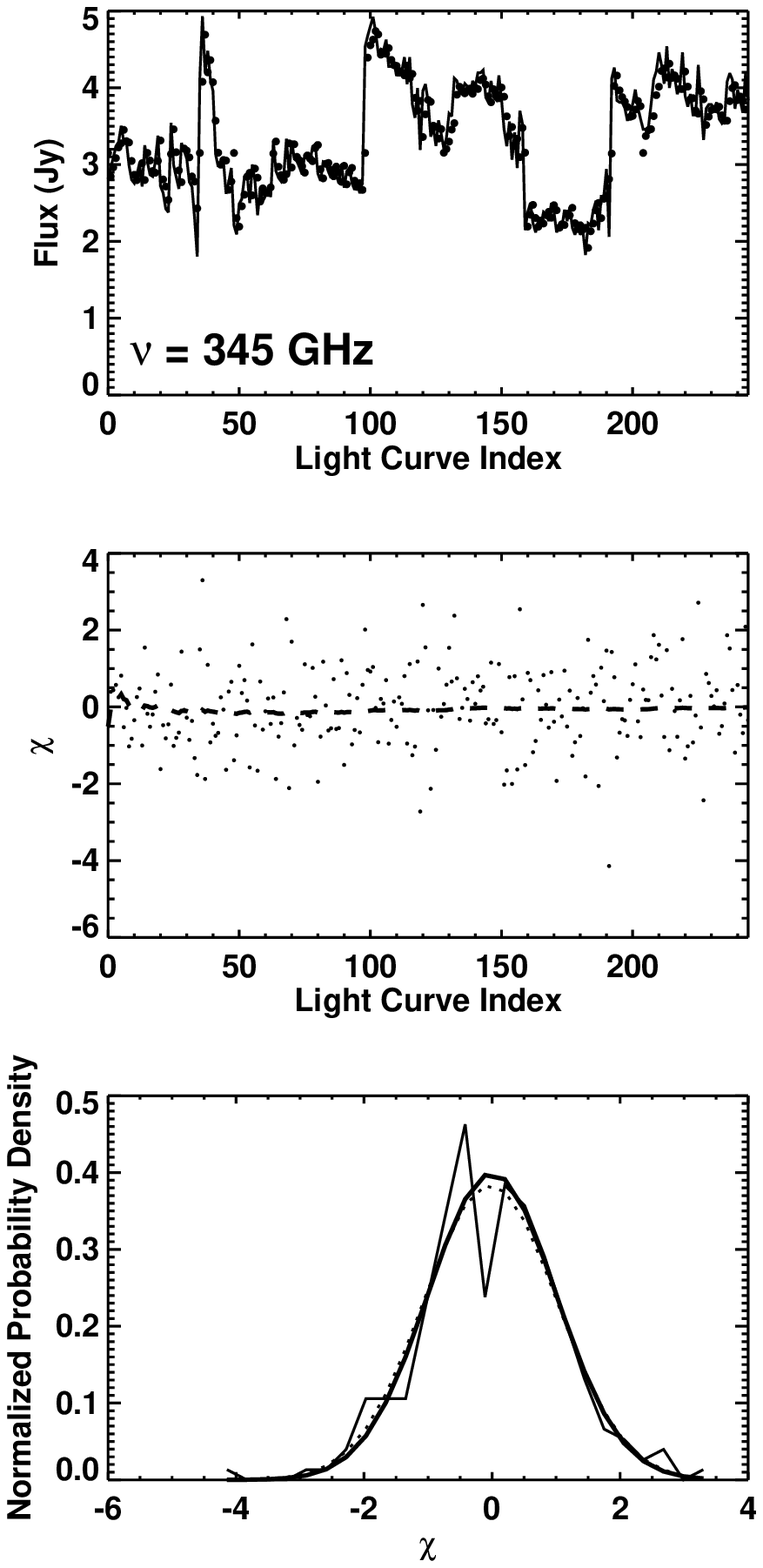}&
\includegraphics[scale=.56]{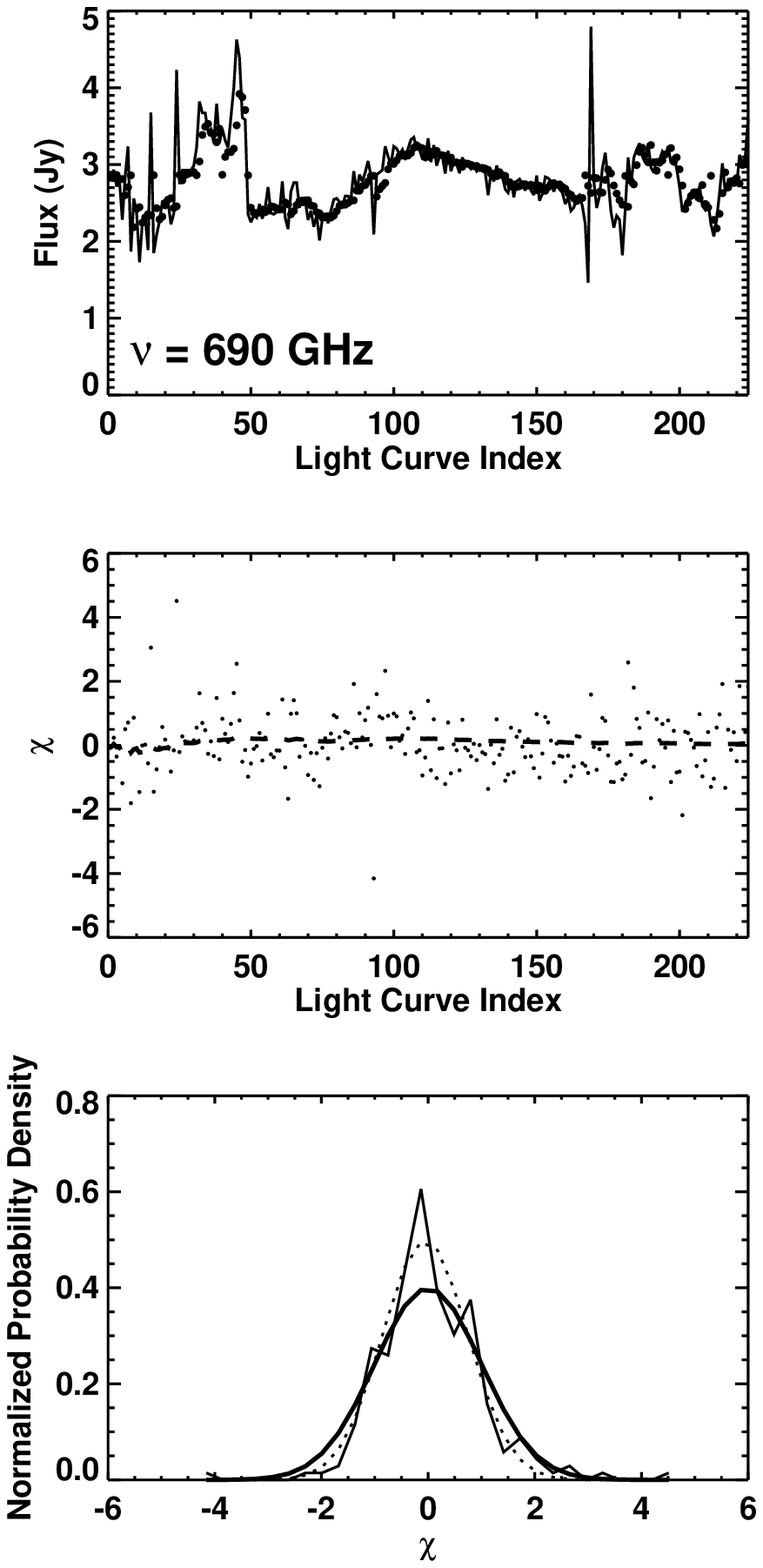}
\end{tabular}
\end{center}
\caption{\label{fit230}Light curve (top panel) as a function of index (solid line) along with the best-fitting damped random walk model (points). The residuals are plotted vs. light curve index in the middle panel (points), along with their running average (dashed line). The bottom panel shows a histogram of the residuals (thin line), a Gaussian fit to the residuals (dotted line), and the expected Gaussian with zero mean and variance of unity (thick line). The three columns correspond to the three observed light curves.}
\end{figure*}

As noted above, we expect the residuals, $\chi$, of the fit to be normally distributed (Gaussian with mean of zero and variance equal to one). Figure \ref{fit230} shows the light curves, CAR(1) fits, residuals and a histogram of the residuals at each frequency. At $230$ and $690$ GHz, the variance of the residuals is smaller than expected. Furthermore, the distribution is still narrower than expected even when uniformly scaling the observational errors to force the overall reduced $\chi^2=1$. 

These results indicate the presence of power in the light curve that is not well described by the DRW model. An excess of small residuals could indicate excess correlation on the shortest time intervals or a power spectrum slope steeper than that of red noise ($\beta=2$, consistent with the allowed range of power law slopes found in \S \ref{lightcurves} ($1.7-3.1$). However, it is also possible that the excess residuals are caused by differences in shapes between the true and model power spectra rather than the slope.

The main goal of the CAR(1) analysis is to measure the characteristic time $\tau$ of the light curves. Therefore, the important issue is whether the difference between the intrinsic and model power spectra lead to biases in our parameter estimates. Qualitatively, the CAR(1) process measures $\tau$ as the time-scale beyond which the light curve is no longer correlated. So as long as the light curve is approximately white noise down to times  $\simeq \tau$, a steepening on shorter intervals should not significantly bias the results. 

To test for bias, we try fitting the light curve excluding time separations $\Delta t < 2$ hours. The estimates of $\tau$ and $\hat{\sigma}$ from this fit is nearly identical to that including all the data, except that the resulting probability distributions are broader. This broadening could be a result of bias, but the best-fitting values of $\tau$ and $\hat{\sigma}$ are nearly identical between the two fits. Further, the broadening of the distributions is fully consistent with the large number of data points excluded ($90\%$). For example, assuming a Gaussian distribution in $\log \tau$, the $1 \sigma$ range should have $\sigma_1 / \sqrt{N_1} = \sigma_2 / \sqrt{N_2}$, where $\sigma_{1,2}$ and $N_{1,2}$ are the confidence intervals and number of data points in fits with different numbers of data points. The expected broadening can be explained by our values of $N_{1,2} = 87, 1052$ between the two fits. We check this hypothesis by randomly selecting $N_1$ points from the full light curve and re-doing the fit. The same residuals at small $\chi$ appear as from using the full light curve, while the allowed range in $\tau$ becomes nearly identical to that when excluding points separated by $\Delta t < 2$ hours. We conclude that the allowed parameter ranges broaden simply from the exclusion of a large fraction of the data. Since we find no evidence of bias, we use the parameter constraints from fitting the full light curves, despite the presence of excess residuals at small $\chi$. We further note that the $345$ GHz light curve fit shows no similar residuals and leads to consistent values of $\tau$. Table \ref{params} provides the parameter constraints from fitting all light curves with log or linear flux density, and with full light curves or excluding short time separations for the $230$ GHz light curve. Confidence intervals are listed in $1-3 \sigma$ as would be expected for a normal distribution.

Submillimetre light curves of Sgr A* are difficult to obtain observationally, and the measured flux density values can be biased by systematic effects such as calibration errors and, particularly in the case of single-dish measurements, confusion with the surrounding emission. To simulate the effects of poor absolute (total flux) calibration and confusion, we add random DC offsets to the data from each day separately and redo the fitting procedure. At $230$ GHz, this does not produce any clear systematic change in the parameter estimates. At $690$ GHz, $\tau$ increases. In both cases, the overall model fit becomes worse. Typically atmospheric effects are removed from the data by calibrating against a known non-variable source, and in some observations the calibration source undergoes atmosphere-induced secular variability on hour time-scales. We simulate systematic errors from inaccurate intraday calibration by adding sinusoidal variations within each day of the light curves. At $230$ GHz, this change decreases $\tau$. It also increases the excess residuals at small $\chi$. It is then possible that instead of an intrinsically steep power spectrum ($\beta > 2$), the observed excess correlations on short times are caused by systematic errors in the intraday light curves. In this case we might expect that removing the systematics could somewhat increase the measured $\tau$ value.

We also explore systematic differences between telescopes. In particular, most of the 230 GHz data come from SMA and CARMA. Fitting these light curves separately, the inferred parameters from the CARMA and SMA light curves are consistent at the $\lesssim 1 \sigma$ level. Fitting separately to data from the other telescopes taken from previously published work gives parameters that are consistent with both of them. Therefore there is no evidence for systematic errors between telescopes.

\begin{figure}
\includegraphics[scale=.6]{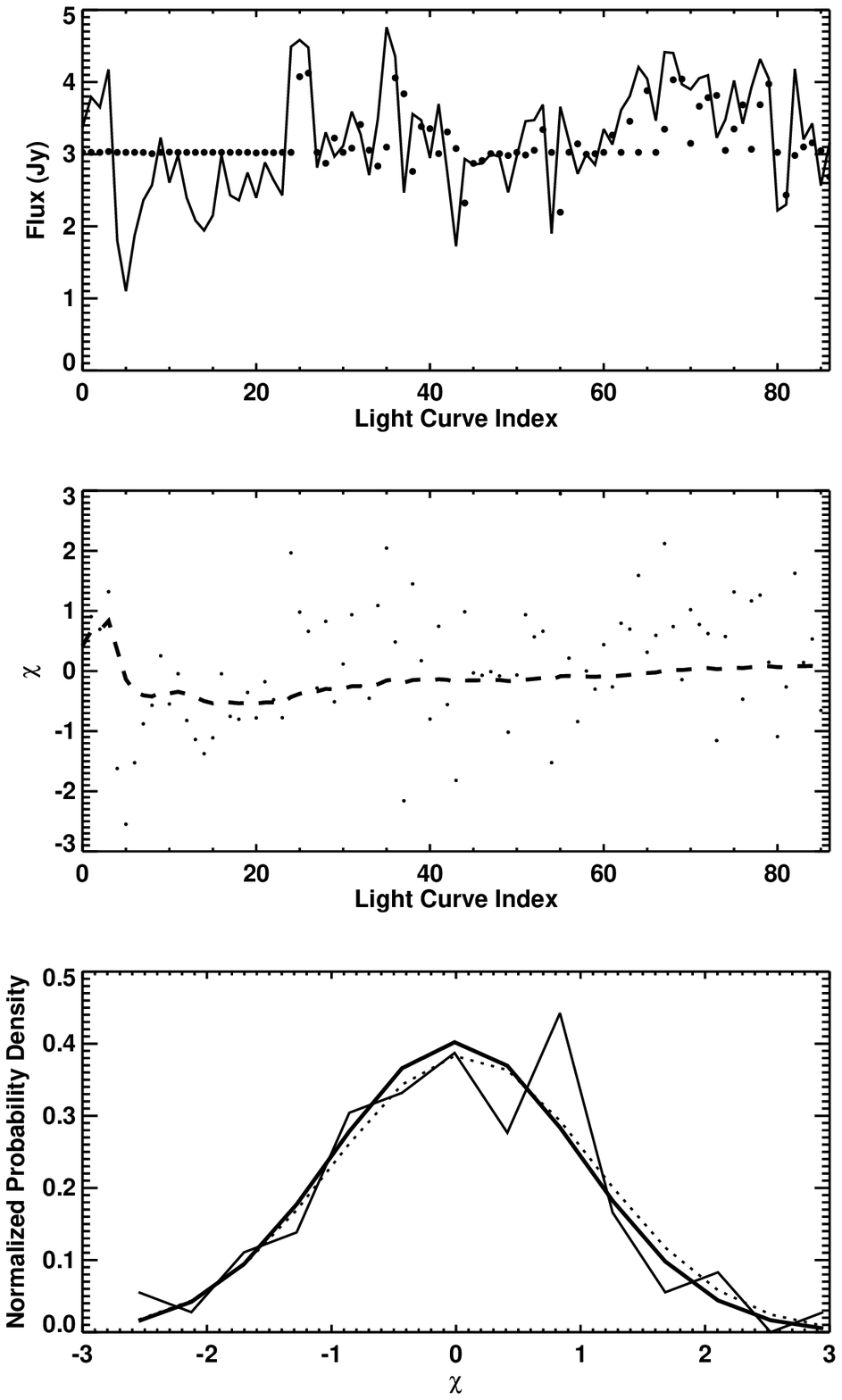}
\caption{\label{fitdt4}As in Figure \ref{fit230}, but excluding data separated by $\Delta t < 2 \rm hr$. The DRW fit is improved over using the full light curve, as discussed in the main text.}
\end{figure}

\section{Implications}
\label{discuss}

We have analysed long, irregularly sampled submm light curves of Sgr A*. From Monte Carlo simulations accounting for sampling biases, we find that the intraday power spectrum can be well fit by a power law with slope $2.3^{+0.8}_{-0.6}$ at $95\%$ confidence, consistent with a red noise process. Since the average submm flux density of Sgr A* has been stable over the full length of the light curves, we expect that the light curve should transition to white noise. We tentatively identify this transition time as a break in the structure function at roughly $\sim 1-100$ hours. However, structure functions calculated in this way are prone to significant biases, including spurious break features. In order to confirm the presence of this transition and to quantitatively measure the associated time-scale without bias from irregular sampling, we model the light curves as a damped random walk (DRW). From this procedure, we measure a characteristic time-scale of $\tau=8_{-4}^{+3}$ hours in the $230$ GHz light curve ($95\%$ confidence), where the constraints are strongest due to the large number of data points. Slightly smaller but consistent values of $\tau$ are inferred at $345$ and $690$ GHz. The analysis is robust to whether we model the flux density or its logarithm---we are unable to distinguish between linear and non-linear noise processes. Further, we find excess residuals that are likely due to either a steep intrinsic power spectrum ($\beta > 2$) at times shorter than $\tau$ ($\lesssim 2$ hours), or to residual systematic intraday errors in the light curves. There is no evidence for bias in the inferred parameters.

The size of the emission region at $230$ GHz has been measured by VLBI observations \citep{doeleman2008,fishetal2011} to be $\simeq 4 R_s$, where $R_s = 2 G M / c^2$ is the Schwarzschild radius, assuming the mass of Sgr A* measured from stellar orbits, $M=4 \pm 0.5 \times 10^6 M_\odot$ \citep[e.g.,][]{ghezetal2008,gillessen09}. However, the inferred size is model-dependent and may be influenced by strong relativistic effects. Geometric \citep{brodericketal2011,binkamruddindexter2013} and physical \citep{broderick2009,brodericketal2011,moscibrodzka2009,dexter2009,dexteretal2010,dexterfragile2013} models which fit the data imply submm emission radii between $2-5 R_s$. These submm emission radii are consistent with the results of previous accretion flow \citep{yuanquataert2003} and jet \citep{falckemarkoff2000} models. Assuming a fiducial emission radius of $3 R_s$, we can compare our measured time-scale for the light curves to the expected orbital ($t_{\rm orb}$) and viscous ($t_{\rm visc}$) times:

\begin{align}
t_{\rm orb} &\approx 0.5 \left(\frac{R}{3 R_s}\right)^{3/2} \rm hr,\\
t_{\rm visc} &\approx 8 \left(\frac{R}{3 R_s}\right)^{3/2} \left(\frac{\alpha}{0.03}\right)^{-1} \left(\frac{H/R}{0.6}\right)^{-2} \rm hr,
\end{align}

\noindent where $\alpha$ is the dimensionless viscosity parameter in standard accretion theory \citep{shaksun1973} and $H/R$ is the disc scale height. Our measurement of $\tau \sim 8$ hours is a factor of $\sim 20$ longer than the orbital time. This suggests that the variability is not due to orbital modulation, e.g., from inhomogeneities in the accretion flow. Eight hours is consistent with the viscous time at this radius, assuming values of $\alpha \simeq 0.03-0.05$ found in current simulations of magnetic turbulence \citep[e.g.,][]{hawleyetal2011,narayanetal2012}. Alternatively, magnetic field reversals occur at intervals of  $\sim 10 t_{\rm orb}$ from dynamo action in several local simulations of MRI driven turbulence \citep[e.g.,][]{guangammie2011}. In this case the reversal time is related to the strength of magnetic stresses driven by the MRI (e.g., $\alpha$). 

In a thick accretion disc the variability does not need to be produced locally but instead can propagate down to smaller radii. A general result of models of time-variable accretion discs is that an $\sim f^{-1}$ power spectrum is expected on the range of local times for radii producing fluctuations \citep{lyubarskii1997}. A break in the power spectrum to white noise is expected to correspond to the relevant time-scale at the outer radius producing fluctuations, while a further steepening of the power spectrum may occur below the shortest times associated with the inner emission radius \citep{churazovetal2001}. If we assume that instead of a local process, the submm Sgr A* variability is due to propagating fluctuations in from larger radius generated on the dynamical time, then the associated maximum outer radius would be $R \sim 20-50 R_S$. This could correspond to a circularization or other transitional radius in the accretion flow. For fluctuations generated on the inflow time, the outer radius is $R \lesssim 10 R_s$ and the variability must be generated close to the submm emission region.

A natural inner scale for the Sgr A* submm emission would be the innermost stable circular orbit of the black hole. Steepening has been seen in power spectra from time-dependent numerical models of Sgr A* \citep{dolenceetal2012,dexterfragile2013}, where it typically occurs on the orbital time of the ISCO ($\simeq 10 \hspace{2pt} \rm min$ for those simulations with high spin). The light curve we have can accurately probe the variability properties on times $\gtrsim 10 \rm min$, since the shortest sampling intervals in the light curve are typically $\gtrsim 5 \rm min$. The combined power spectrum from all individual epochs from minutes to $\simeq 3$ hours is well fit by a single power law model with a red noise or steeper slope. There is no sharp cutoff down to the shortest time intervals probed, suggesting that the submm emission persists down close to the ISCO, if not the event horizon, even if the variability is produced on the local dynamical time. 

A previous analysis of much longer wavelength ($1.3 \rm cm$) radio light curves found two characteristic time-scales, at $\tau > 1$, $> 10$ days \citep{macquartbower2006}. The day time-scale was argued to be intrinsic to Sgr A*, while the longer one was associated with scintillation effects along the line of sight. This radio value of $\tau$ is somewhat larger than our submm result, as expected if the self-absorbed radio emission arises in a photosphere much farther from the black hole as measured by VLBI observations \citep[e.g.,][]{bower2006}.

\citet{mauerhan2005} found excess power at a time $\simeq 2.5$ hours in the periodogram of a $3$ mm light curve compared to a noise process with $P(f) \propto f^{-1}$. This result is consistent with the emission size from VLBI observations \citep[e.g.,][]{krichbaumetal1998} only if the variability was produced on the local orbital time. This is a significantly shorter time-scale from emission arising farther from the black hole than we find here. It is not clear whether the excess inferred by \citet{mauerhan2005} is due to a break in the power spectrum, and if so whether the break is to white noise or steeper red noise. If the time-scales are equivalent, that would indicate a different variability mechanism operating at $3$ mm than at $1.3$ mm. While surprising, this is not implausible given that spectral fits to Sgr A* suggest that different electron populations could produce the submm and mm emission \citep{yuanquataert2003}, and these electrons could come from different dominant dynamical components as well \citep{yuanetal2002}. If the \citet{mauerhan2005} excess is instead interpreted as a QPO, then we do not see any comparable excess of power at the $99.7\%$ level. At that frequency, our $230$ GHz light curve would be sensitive to an oscillation with an amplitude of $\gtrsim 0.3$ Jy.

A power spectrum measured from a long near-infrared light curve showed a break in the power spectrum, also at $\simeq 2.5$ hours \citep{meyeretal2009}. The $\tau$ value measured from the $230$ GHz light curve is larger than this value at $> 3 \sigma$ significance. It is likely that the same is true at $345$ and $690$ GHz, although the constraints are worse due to limited data at those frequencies, lowering the significance. The post-break power spectrum slope they found ($\beta = 2.1^{+0.8}_{-0.5}$) is consistent with previous work in the NIR \citep{doetal2009} and with our measured $230$ GHz power spectrum slope. 

If the NIR and submm variability are driven by the same underlying local process with the expected $t \propto R^{-3/2}$  dependence, our measured $\tau$ value implies that the NIR emission radius is $\approx 1/6$ that of the submm. For the assumed submm emission radius of $3 R_s$, the NIR emission is coincident with the ISCO even for a maximum spin of $0.998$. In reality, the maximum spin is likely lower for geometrically thick accretion flows like in Sgr A* \citep{gammieetal2004,mckinneyetal2012}. It therefore seems unlikely that the submm and NIR variability arise from the same process operating at different radii if  $R \simeq 3 R_s$ is really the radius associated with our measured submm variability time-scale. 

Alternatively, if the mm-VLBI is measuring a small region of the accretion flow at larger radius, the variability could correspond to a distance as large as $\simeq 25 R_S$ if the fluctuations are generated on the orbital time. In this case, the NIR emission could be produced at $\simeq 8 R_s$ by the same mechanism. However, the size measured by mm-VLBI has been stable over 4 epochs \citep{fishetal2011}, even as the total flux density of the compact component has changed by $\simeq 1$ Jy, which seems unlikely for a small part of the accretion flow at larger radius. 

Connections between submm/NIR variability have been suggested by observations where peaks in the submm light curves (``flares'') appear delayed from those in the NIR by $\sim 2$ hours \citep[e.g.,][]{eckart2008sim,marrone2008}: the NIR light curve is modeled as emission from adiabatically expanding  plasma, which cools and produces submm emission at a larger size. Qualitatively this could explain the difference between the characteristic NIR/submm time-scales. These models predict a change in mm-VLBI size during flares which depends on the relative strength of the flare \citep{huangetal2012}, which was not detected by \citet{fishetal2011} during a $\simeq 20\%$ increase in total flux density. \citet{eckartetal2012} proposed that the submm variability could be a combination of flares from adiabatically expanding gas and lower level fluctuations from turbulence, so that the observed size would change in some cases but not others. Our light curves show no clear evidence for multiple processes: the flares are just larger excursions expected from a DRW. Fitting instead with a mixture of DRW processes \citep{kellyetal2011} does not significantly improve the fit. This model can be tested with future simultaneous mm-VLBI and NIR/X-ray observations.

Other work has proposed that the submm and NIR emission are anti-correlated \citep{hauboisetal2012}, perhaps as a result of decreased magnetic field strength following strong reconnection events \citep{doddsedenetal2010}. The significant difference in characteristic $\tau$ values for the submm/NIR could potentially be explained by different physical time-scales for triggering reconnection events and for subsequently increasing  the magnetic field strength. 

This CAR(1) method has been used to estimate $\tau$ in optical quasar light curves  \citep{kellyetal2009,macleod2010,kozlowskietal2010}. In particular, \citet{kellyetal2009} measured a roughly linear trend with black hole mass, $\tau \approx 80 \hspace{2pt} (M_{\rm bh}/10^8 M_\odot) \hspace{2pt} \rm days$. This relation predicts a $\tau$ value of a few days for $M_{\rm Sgr A*} \simeq 4\times10^{6} M_\odot$, an order of magnitude larger than found here. The size of the optical emission region has been measured in several sources using gravitational microlensing \citep[e.g.,][]{morganetal2010}, with typical sizes $\simeq 30-50 R_S$ \citep[e.g.,][]{daietal2010}. Naively using the expected $R^{3/2}$ scaling roughly leads to agreement between with their relation. This may just be a coincidence: for $H/R \sim 1$ expected for radiatively inefficient sources, the thermal and viscous times become identical. The $\tau$ values found in AGN are inconsistent with a viscous origin \citep{kellyetal2009} given the $\propto (H/R)^{-2}$ and the thin disks expected at luminosities comparable to Eddington. The emission radius in both cases is uncertain at the factor of a few level, which leads to order of magnitude uncertainty when comparing the time-scales.

X-ray variability studies of stellar mass black hole binaries and AGN have also identified power law breaks, and have argued for a linear scaling with black hole mass \citep{uttleymchardy2005} which becomes significantly better when including a correction for bolometric luminosity \citep{mchardyetal2006}. Both the NIR Sgr A*  \citep{meyeretal2009} and our submm time-scale are many orders of magnitude larger than expected from this relation, because of the extremely low luminosity of Sgr A*. Ignoring the correction, both $\tau$ values are within the range seen from X-ray light curves of AGN, scaled to the mass of Sgr A*. Again it is not clear whether this is meaningful, especially given the few order of magnitude scatter in the X-ray sources. Physically we might expect the NIR/X-ray, rather than the submm, emission in Sgr A* to correspond more closely to the X-ray emission in AGN, but we cannot determine this based on scaling the variability properties alone.

As well as comparisons to characteristic time-scales in the accretion flow and to those measured in other wavebands and/or sources, the quantitative description of the submm variability presented here will be useful in the near future. Theoretical light curves from numerical simulations \citep[e.g.,][]{dexter2009,dexteretal2010,dexterfragile2013,dolenceetal2012,shcherbakovmckinney2013}, adiabatic expansion, or RIAF+hotspot models \citep{doelemanetal2009} can be compared directly with these results by considering the probability that any given theoretical light curve would arise from a viable DRW process. In particular, using numerical GRMHD simulations with an aligned accretion torus and black hole spin axis, \citet{dexter2009} and \citet{dolenceetal2012} found relatively steep power law slopes ($\beta \simeq 2-3$) at $230$ GHz on time intervals longer than $5-10$ minutes, consistent with our inferred range. The steep power law slope we find is more compatible with the low (face-on) rather than high inclination model of \citet{dexter2009}, but the slope may depend on the initial conditions or other specifics of the simulation used there. \citet{dexterfragile2013} found a much shallower power slope ($\beta \simeq 1$) from a simulation where the black hole and accretion flow were misaligned, inconsistent with our measured value. \citet{shcherbakovmckinney2013} predicted the presence of a QPO with a $35$ min period and an amplitude $\simeq 0.15$ Jy at $230$ GHz, marginally inconsistent with our limit of $0.1$ Jy for that period. These results suggest that quantitative variability properties constitute a new and important means to constrain theoretical models of Sgr A*.

Similarly, the arrival of the few earth mass gas cloud G2 \citep{gillessenetal2012} at pericentre in late 2013 \citep{gillessenetal2013} or early 2014 \citep{phiferetal2013} could trigger additional accretion and thus a secular increase in the submm flux density of Sgr A*. Future submm light curves of Sgr A* can be compared to our results to determine the significance of changes in the flux density of Sgr A*, and to quantify any resulting increases in the accretion rate.

\section*{acknowledgements}
JD thanks E. Churazov, T. Do, C. Gammie, M. Kerr, and E. Quataert for useful discussions. M. Gurwell provided assistance with SMA data calibration. This work made use of an IDL routine written by James Davenport. Support for CARMA construction was derived from the states of California, Illinois, and Maryland, the James S. McDonnell Foundation, the Gordon and Betty Moore Foundation, the Kenneth T. and Eileen L. Norris Foundation, the University of Chicago, the Associates of the California Institute of Technology, and the National Science Foundation. Ongoing CARMA development and operations are supported by the National Science Foundation under a cooperative agreement, and by the CARMA partner universities.
\footnotesize{
\bibliographystyle{mn2e}
\bibliography{master}
}
\label{lastpage}

\end{document}